\documentclass[aoas,MSNbibl,nameyear,dvips]{arximspdf}
\usepackage{dcolumn}
\usepackage{graphicx}
\usepackage{url,breakurl}

\doi{10.1214/14-AOAS741} 
\volume{8}
\issue{3}
\pubyear{2014}
\firstpage{1583}
\lastpage{1611}

\makeatletter
\newcolumntype{d}[1]{D{.}{.}{#1}}

\newcommand{\rright}{\right}
\newcommand{\lleft}{\left}
\makeatother

\begin{document}
\begin{frontmatter}

\title{A Bayesian approach for predicting the popularity of tweets}
\runtitle{Predicting popularity of tweets}

\begin{aug}
\author[A]{\fnms{Tauhid}~\snm{Zaman}\corref{}\ead[label=e1]{zlisto@mit.edu}},
\author[B]{\fnms{Emily B.}~\snm{Fox}\ead[label=e2]{ebfox@stat.washington.edu}}
\and
\author[C]{\fnms{Eric T.}~\snm{Bradlow}\ead[label=e3]{ebradlow@wharton.upenn.edu}}
\runauthor{T. Zaman, E.~B. Fox and E.~T. Bradlow}
\affiliation{Massachusetts Institute of Technology,
University of Washington\\
and
University of Pennsylvania}
\address[A]{T. Zaman\\
Sloan School of Management\\
Massachusetts Institute of Technology\\
Cambridge, Massachusetts 02139\\
USA\\
\printead{e1}}
\address[B]{E.~B. Fox\\
Department of Statistics\\
University of Washington\\
Box 354322\\
Seattle, Washington 98195\\
USA\\
\printead{e2}}
\address[C]{E.~T. Bradlow\\
The Wharton School\\
University of Pennsylvania\\
Philadelphia, Pennsylvania 19104\\
USA\\
\printead{e3}}
\end{aug}

\received{\smonth{4} \syear{2013}}
\revised{\smonth{1} \syear{2014}}

%
\begin{abstract}
We predict the popularity of short messages called
\emph{tweets} created
in the micro-blogging site known as Twitter.
We measure the popularity of a tweet by the time-series path of its
\emph{retweets}, which is when people forward the tweet
to others. We develop a probabilistic model for the evolution
of the retweets using a Bayesian approach, and form predictions
using only observations
on the retweet times and the local network or ``graph''
structure of the retweeters.
We obtain good step ahead forecasts and predictions of the
final total number of retweets even when only a small fraction
(i.e., less than one tenth) of the retweet path is observed.
This translates to good predictions
within a few minutes of a tweet being posted, and has potential
implications for understanding the spread of broader ideas, memes
or trends in social networks.
\end{abstract}

%
\end{frontmatter}

\section{Introduction}
The rapid rise in the popularity of online
social networks has resulted in an explosion
of user-generated content. There is a wide
variety in the type of content---it can be a user
comment, a photograph, a movie or a link to a news
article. Typically, in these online social networks,
users form connections with other users, producing a
social graph. For example, in the micro-blogging site Twitter,
these connections are known as \emph{followers} and the
resulting social graph is known as the \emph{follower graph}.
When a user generates a piece of content, it becomes
visible to all of his or her
followers in the social graph. The content spreads
through the social graph if these followers subsequently
repost the content so their followers can see it and potentially repost
it further.

In this work we focus on the micro-blogging site Twitter
which has over 230 million active users as of November 2013
[\citet{reftwitterSEC}].
The user-generated content in Twitter is composed of short messages
known as \emph{tweets}
containing up to 140 characters,
which can also contain images or links to news articles or videos.
Tweets are spread
through the Twitter follower graph by the act of \emph{retweeting}, which
is when a user forwards a tweet to his or her followers.

Our goal in this work is to predict the popularity of a tweet
by predicting the time path of retweets it receives. We aim
to make these predictions very early on in the lifetime of
the tweet, sometimes within minutes of it being posted.
We use a Bayesian model to describe the evolution
of the retweets of a tweet. With this model we make
predictions for the total number of retweets a tweet
will receive using information from early retweet
times, the retweets of other tweets and summaries of the follower graphs.

The remainder of the paper is organized as follows. In Section~\ref{secprevwork}
we describe related work. In Section~\ref{secdataOverview} we
provide a description of the
data utilized and an exploratory set of analyses of it that
guide the proposed probabilistic model
of Section~\ref{secmodel}. We present our posterior computations
via Markov chain Monte Carlo (MCMC) in Section~\ref{secposterior}.
In Section~\ref{secresults} we present an analysis of our model's
predictive performance
on our Twitter data, including a comparison to benchmark models from the
extant literature and nested versions of our model. We discuss
extensions to this
research in Section~\ref{secextensions}.

\subsection{Previous work}\label{secprevwork}

There has been much recent interest in the retweet prediction
problem, albeit in terms of a slightly different type of prediction task.
In particular, recent extant research [\citet{refzaman,reftwitterTree}]
tried to predict the existence of a
retweet between a particular pair of users. While this is an important
problem in graph formation or viral spreading across vertices, it is
a notably different problem than
addressed here due to the precision and pairwise specificity required.

\citet{refsuh} used a generalized linear model
to understand what features influenced the chance of a tweet being retweeted
by anyone. Other work [\citet{refhong,refbandari}] built upon
this and used a variety of algorithms to try to predict not the exact
number of retweets, but rather a coarse
interval for the number of retweets of a tweet. Similar techniques were used
by \citet{refnaveed} and \citet{refpetrovic} to predict the
probability that a tweet receives any retweets, which by definition
is nested within the problem we consider.

In contrast to these previous works, we aim to predict
the entire time path, and hence the eventual number of retweets
of a tweet. This is similar to
\citet{refszabo} who use a linear model
to predict the popularity of
stories on \href{http://www.Digg.com}{Digg.com} and videos on YouTube after 30 days by observing
their popularity after one hour and one week, respectively.
Other related work is
\citet{refagarwal} who attempt to make one-step
ahead predictions of the
click-through rates of online news stories
with a spatial--temporal model that utilizes
the time-varying click-through rate of an article along
with its spatial position on a webpage. The problem
of predicting the structure of time evolving citation networks
is studied in \citet{refego}.
Our prediction
goal is similar to these works, but as we demonstrate in Section~\ref{secresults},
our approach produces accurate predictions for the final number of retweets
using only minutes of observations, rather than hours or days.
Given the Bayesian approach utilized here, accurate predictions
are possible for a given tweet's retweet
path even when there are no available data other
than that of other retweet paths observed so far, especially
if one utilizes covariates describing the tweets, retweets
and their authors (an area for future research).

\section{Data overview}\label{secdataOverview}
In this section we describe the retweet data we obtained
and present exploratory data analysis
of some basic features. This analysis is useful
in providing an understanding of the scales
associated with the data (number of retweets of a typical tweet,
time-scale over which a typical tweet is retweeted) and in guiding
our more formal modeling choices.

\subsection{Data description}
We collected retweet data that cover a fairly wide
array of topics and also have a wide range of retweet graph sizes.
The topics include music, politics and
miscellaneous everyday events.
Our data set consists of 52 different tweets which were
selected through manual exploration of Twitter and are available
in the supplemental materials [\citet{refsuppA}]. We refer
to these original tweets as \emph{root tweets}. For each root
tweet, we used the Twitter Search API [\citet{reftwitterSearchAPI}]
to find all retweets. We used root tweets which were at least a
week old to make sure that there were likely to be no more retweets occurring.
The search API provided us with the retweet times and identity
of the users who retweeted. Also, since the Search API could
only return a maximum of 1800 results, we did not
look at root tweets with more than this many retweets.
Based on previous empirical studies [\citet{refiran,refmiacha}],
this maximum number of retweets covers a large fraction of tweets in Twitter
and does not represent a significant limitation. However, it is an open
research question as to what degree the empirical patterns we observe
will hold for tweets with a large number of retweets.

From the text
of the retweet, we are able to identify the person
that the user retweeted (the username
following the text ``RT@''). For example,
if user Alice posted the tweet ``Hello'' and user Bob
retweeted this root tweet, it would appear as
``RT@ Alice: Hello.'' We then used the Twitter API
to find the number of
followers of the root user and each user who retweeted it. The
number of followers
will act as a covariate in our predictive model. In
particular, the number of followers for a given user
represents both the potential retweet base for a given
tweet and also a significant moderator of the
speed and timing of retweets.

We associate with each root tweet a directed
\emph{retweet graph}. We will utilize the
following notation for the different data associated with
the retweet graph. We denote the root tweet as $x$ which
is tweeted by root user $v_0^x$. The retweet graph associated with $x$
which we observe at time $t$ is denoted
$G^x(t)=(V^x(t),E^x(t))$. The vertex set $V^x(t)$ includes the root
user (who
tweets at $t=0$) and
all users who retweet the root tweet before time $t$.
A directed edge $(u,v)\in E^x(t)$ points
from user $u$ to user $v$
if $v$ retweets $u$ before $t$. We will denote
the total number of retweets in $G^x(t)$ by $m^x(t)=|V^x(t)|-1$.
We define the final number of retweets of $x$ as
$\lim_{t\rightarrow\infty}m^x(t)=M^x$
and it is the arrival of retweets and attained
$M^x$ that we wish to predict.

\begin{figure}

\includegraphics{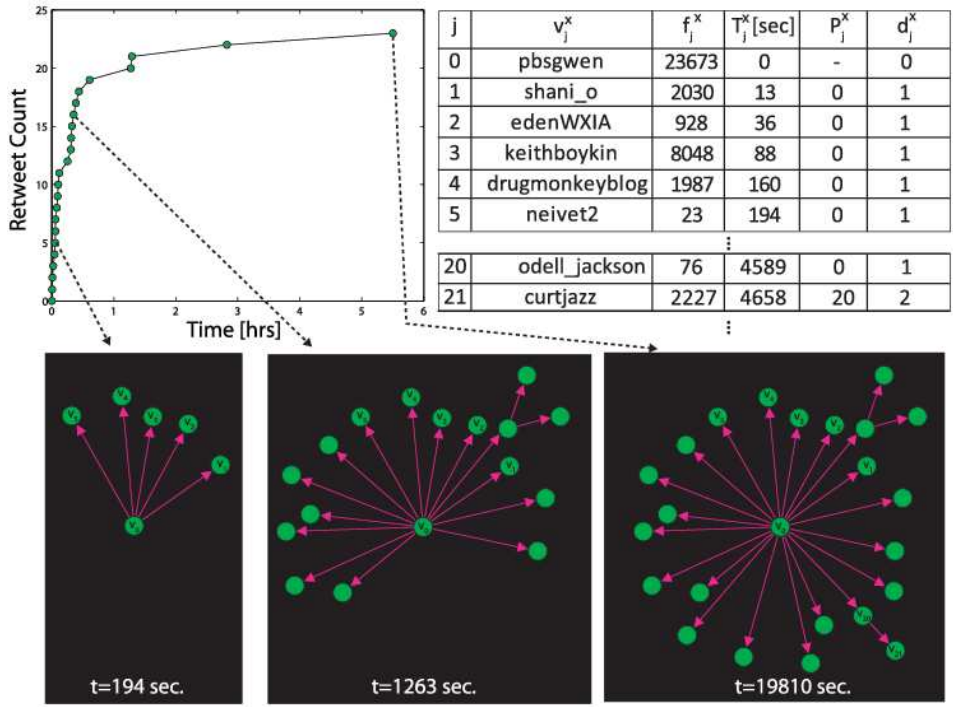}

\caption{Data for the root tweet ``Cory Booker has
never worked a day in his life. Not. \#corybookerstories'' by root user
pbsgwen. The table shows the relevant data for the retweet graph
for several users. The plot shows the number of retweets of the root tweet
versus time. Images of the retweet graph at different times
are also shown.}\label{figrtgraph1}
\end{figure}

We will index the users in the retweet graph with the variable $j$.
The root user is indexed by $j=0$, and all other users have $j>0$.
User $j$ who retweets $x$ is denoted by $v_j^x $ for $j=1,2,3,\ldots.$ The
time of
this user's retweet is denoted $T_j^x $, with $T_0^x=0$ (the root
tweet occurs at time $0$). User $v_j^x $ has $f_j^x $ Twitter followers
and is $d_j^x $ ``hops'' from the root user $v_0^x$ in the retweet
graph. The parent\vspace*{1pt}
of $v_j^x $ in the retweet graph is denoted $P_j^x$.
To illustrate these definitions, we show in Figure~\ref{figrtgraph1}
an example of the retweet graph for a root tweet.
Included are pictures of the evolution of the retweet graph, a plot
of the number of retweets versus time and a table showing the
aforementioned summary data for several users in the retweet graph.
As we can see, this particular root tweet has almost all
of its retweets at depth one (one hop from the source), which is a common
pattern for our data set as discussed below.

\subsection{Size, lifetime and depth of retweet graphs}\label{secdataAnalysis}
We first look at the size and lifetime of the 52 retweet graphs.
The root tweets we collected had between 21 and 1260 retweets. The time
for the final retweet to occur ranged from a few hours to a few days as
some of the final retweets had very large retweet times. A more
stable measure of the lifetime of a root tweet is the time to reach
50\% (the
median) of its total retweet count. The median retweet
times ranged from four minutes to three hours, with most
being less than one hour.

%

%
%
\begin{figure}[b]

\includegraphics{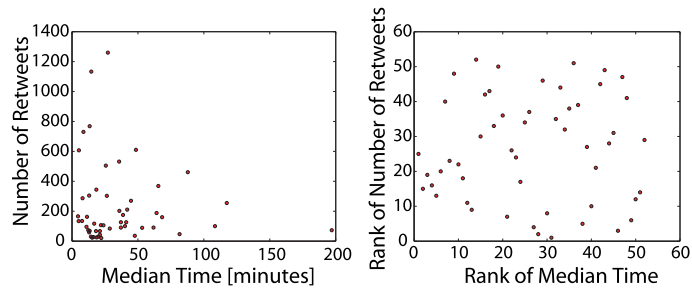}

\caption{(Left) total number of retweets versus median
retweet time for different root tweets.
(Right)~rank of total number of retweets versus
rank of median retweet time for different root tweets.}
\label{figMtrueVsTmedian}
\end{figure}

We plot the total number of
retweets versus the median retweet times for the 52 root tweets in
Figure~\ref{figMtrueVsTmedian}. We also plot the rank of each tweet's
median retweet time versus the rank of its total number of retweets
among our 52 source tweets.
The Pearson correlation coefficient
for the median retweet times and the eventual number of retweets is
$-$0.12 ($p$-value${}=0.49$) and the Kendall tau rank correlation coefficient
is 0.03 ($p$-value${}=0.84$).
Therefore, we do not have evidence to reject the null hypothesis that
the eventual number of retweets is uncorrelated with the median retweet
time. Instead, this suggests the potential value of our model over
purely exploratory approaches. In particular, it is important to model
the retweet interarrivals for our prediction task. Thus, simply predicting
the total number of retweets from the median (or simple central summary)
is unlikely to yield accurate predictions.

%
\begin{figure}

\includegraphics{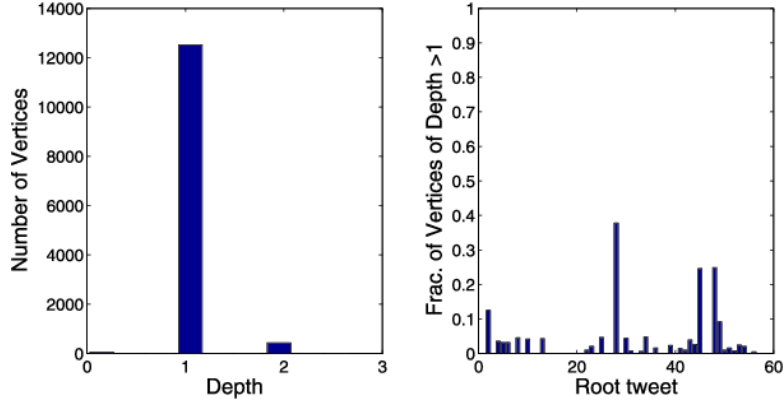}

\caption{Histogram of (left) the fraction of users at different
depths in all 52 retweet graphs and (right)
the fraction of vertices of depth greater than one in the
retweet graph for each root tweet.}
\label{figdepth}
\end{figure}

We next explore the structure of the retweet graphs. In particular,
we look at the number of vertices one hop and
more than one hop from the root user. For the 52 root tweets, there are 11,882
retweeters who are one hop
from the root user and only 314 retweeters more than one hop from the
root user.
Figure~\ref{figdepth} shows the histogram of vertices at different
depths in all of
the retweet graphs, along with a plot of the fraction of vertices more
than one
hop from the root user for each retweet graph. As can be seen, retweet
graphs typically have
most vertices at depth one, but occasionally they have some vertices at
depth greater than one,
suggesting that root tweets get retweeted much more often than the
retweets get retweeted.
This fact agrees with previous studies done on retweet graph
structures [\citet{reftwittermoon,refgoelRetweetDepth}] and is key
to our ability to predict $M^x$ early, even before potential retweets from
those two hops or more are taken into account.
We have found that the follower count of the root user has little
correlation with the retweet graph depth
(Pearson correlation coefficient${}= 0.13$, $p$-value${} = 0.28$).
However, when a retweet graph has depth greater than one, it is
typically due
to a user with a large number of followers.
The median follower count of users in the retweet
graph who are not the source but
get retweeted is~1,142,923.

\subsection{Reaction times}\label{secreaction}
\begin{figure}

\includegraphics{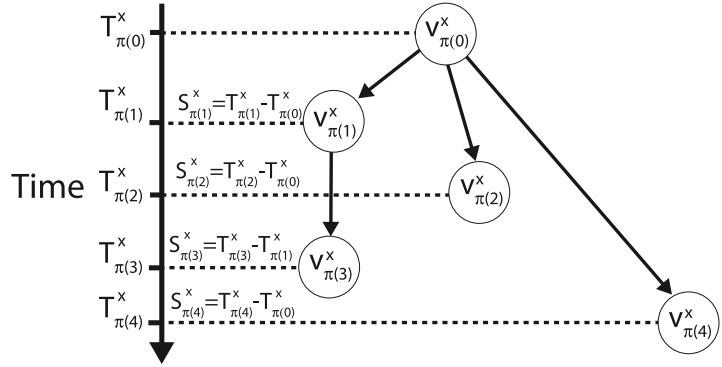}

\caption{Description of reaction times for a retweet graph. The
vertical position of vertices indicate when they retweeted, with time
increasing as one goes down. The reaction time on
each edge is expressed in terms of the retweet times
of the vertices.}\label{figreactionTime}
\end{figure}

Given, as before, that user $v_j^x $ retweets the root tweet at time
$T_j^x $, we define the \emph{reaction time}
$S_j^x =T_j^x -T_{P_j^x}^x$
as the elapsed time between when
the parent of $v_j^x $ (re)tweets and $v_j^x $ retweets.
That is, $S_j^x $ is the time that it takes $v_j^x $ to react
and retweet after the root tweet becomes
visible to $v_j^x $ via its parent's (re)tweet. We define $\pi$
as the permutation that orders the $M^x$ retweet times $T_j^x $
from minimum to maximum. That is,
$T^x_{\pi(0)}\leq T^x_{\pi(1)}\leq\cdots  \leq T^x_{\pi(M^x)}$.
It is\vspace*{1pt} important to note that $\pi$ corresponds to the sequence
in which we observe the retweet times for a root tweet.
Figure~\ref{figreactionTime} provides a graphical
explanation of the reaction times in terms
of retweet times.
%
%

To begin a more formal exploration of our data, we
first consider a simple and non-Bayesian
model in which each $S_j^x $
is assumed to be an i.i.d. log-normal\vspace*{-1pt} random variable with parameters
$\tau^x$ and $\alpha^x$: $\log(S_j^x )\sim\mathcal N(\alpha
^x,(\tau^x)^2)$.
We take the parameters of the log-normal to be different for each root tweet
$x$, but the same for each user within a given retweet graph. This
assumption takes into account the fact that there can
be heterogeneity of these parameters which depends on the content
of the root tweet.

%
\begin{figure}

\includegraphics{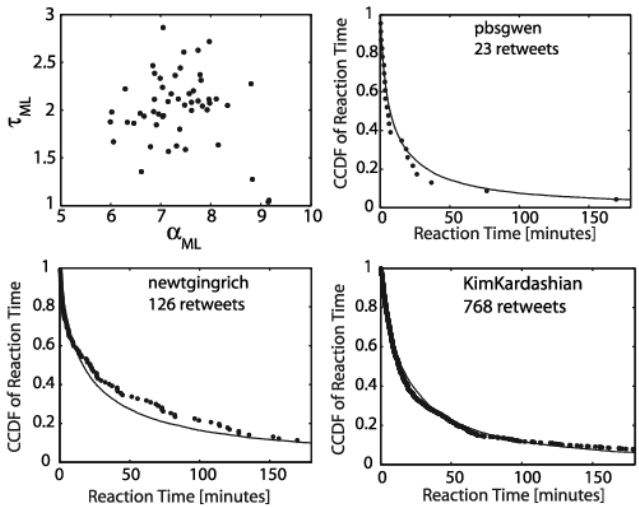}

\caption{(Top left) scatter-plot of ML estimates of $\alpha^x$ and $\tau^x$ for different
root tweets. The remaining figures are plots of the
empirical reaction time complimentary cumulative distribution
function (CCDF) (black circles) and the CCDF of log-normal distributions
using the ML parameter estimates (solid line) for three different root tweets
representing the 2.5 (top right), 50 (bottom left) and
95 (bottom right) percentiles of retweet graph size in our data set.
For each root tweet, we show the root user for the tweet and the number
of retweets in total it received.}\label{figlognormal}
\end{figure}

To assess the log-normal assumption, we calculate the maximum
likelihood~(ML) estimate of $\alpha^x$ and $\tau^x$ for each root
tweet. Given
a set of reaction times $S_j^x $ for $j=1,2,\ldots,M^x$, the ML estimates are
straightforwardly given by
\begin{eqnarray}
\alpha^x_{\mathrm{ML}} &=& \frac{1}{M^x}\sum
_{j=1}^{M^x}\log \bigl(S_j^x
\bigr),\qquad \tau^x_{\mathrm{ML}} = \sqrt{
\frac{1}{M^x}\sum_{j=1}^{M^x} \bigl(\log
\bigl(S_j^x \bigr)-\alpha^x_{\mathrm{ML}}
\bigr)^2}.
\nonumber
\end{eqnarray}
%
In Figure~\ref{figlognormal} (top left) we show a scatter-plot of
$\alpha^x_{\mathrm{ML}}$
and $\tau^x_{\mathrm{ML}}$ for
different root tweets $x$. All parameter values are evaluated
with reaction times measured in seconds. The mean and standard
deviation of $\alpha^x_{\mathrm{ML}}$ is 7.31 and 0.73, respectively.
The mean and standard
deviation of $\tau^x_{\mathrm{ML}}$ is 2.31 and 0.31, respectively, and
we clearly see some heterogeneity over $x$.
To assess fit,
we show in Figure~\ref{figlognormal}
the empirical complimentary cumulative distribution function
(CCDF) of the reaction times along with the
CCDF of a log-normal distribution using the ML estimates for the
parameters for three root tweets representing the 2.5 (small size, top
right), 50 (medium size, lower left) and
95 (large size, lower right) percentiles of retweet graph size in our
data set.
Qualitatively, the log-normal curves provide a reasonable fit for the
reaction times.

The observation of log-normally distributed reaction times
has occurred in other application areas. For instance, \citet{refemail}
observed that the time for people to respond to emails follows
a log-normal distribution. \citet{refbrown} observed that call
durations in call centers
follow a log-normal distribution. In the psychology literature
there have been different models proposed to explain the
origin of log-normal reaction times in different contexts
[\citet{reflog-normal1,reflog-normal2}]. However,
these models do not apply directly to Twitter and it is
interesting to see the same general empirical pattern replicated here.

\subsection{Retweet graph structure}\label{secgraph}
In this section we provide an initial exploration
of the effects of the number of
followers, $f_j^x $, and distance from the root, $d_j^x $,
on the probability of a user's tweet being retweeted.
Once a user $v_j^x $ (re)tweets\vspace*{2pt}
in the retweet graph for a root tweet $x$, the
(re)tweet appears in the Twitter feed (timeline) of
all of $v_j^x $'s followers. Some number of these
followers will subsequently retweet~$v_j^x $. We denote\vspace*{2pt} this number by $M_j^x $, which is equal
to the out-degree of $v_j^x $ in the completed retweet graph
once the root tweet has stopped spreading. We assume that
each of the $f_j^x $ followers of $v_j^x $ will independently retweet
$v_j^x $
with probability $0\leq b_j^x \leq1$. This gives $M_j^x $ a binomial
distribution
$\operatorname{Bi} (f_j^x,b_j^x  )$. We note\vspace*{2pt} that this
assumption of conditional
independence across followers is reasonable because
retweeters are unlikely to be connected to other retweeters
and, hence, there is no ``visibility'' between the $f_j^x $ followers.
In our data set, the average of ratio of cycle forming follower edges
to all possible follower edges is 0.01. This means that
follower edges which connect users in addition to those connected
via retweets represent less than 1\% of all possible
follower edges. For other networks there may be generalizations needed.

We assume the retweet probability $b_j^x $ depends upon two pieces of
information:
the number of followers $f_j^x $ of $v_j^x $ and the distance
$d_j^x $ of $v_j^x $ from $v_0^x$
in the retweet graph. This makes conceptual sense
as these two variables represent the potential retweet base
and the ``degree of closeness'' of each vertex, respectively.
We model $\operatorname{logit}(b_j^x )$ as
\begin{eqnarray}
\operatorname{logit}\bigl(b_j^x \bigr) &=&
\beta_0 + \beta_f\log\bigl(f_j^x
+1\bigr)+\beta_d\log\bigl(d_j^x +1\bigr)+
\varepsilon^x_j\label{eqmub},
\end{eqnarray}
%
where $\varepsilon^x_j\sim\mathcal N(0,\sigma_b^2)$.
For this exploratory analysis (formal model in Section~\ref{secmodel}), for each user
$v_j^x $
we\vspace*{1pt} estimate $b_j^x $ as $\widehat{b}_j^x =M_j^x /f_j^x $.
We then perform a linear regression of $\operatorname{logit}(\widehat
{b}_j^x )$
on $\log(f_j^x +1)$ and $\log(d_j^x +1)$ for all users in all root tweets.
Here, we only include users for which $M_j^x \geq1$ so that
$\operatorname{logit}(\widehat{b}_j^x )$ will be finite.

%
\begin{figure}

\includegraphics{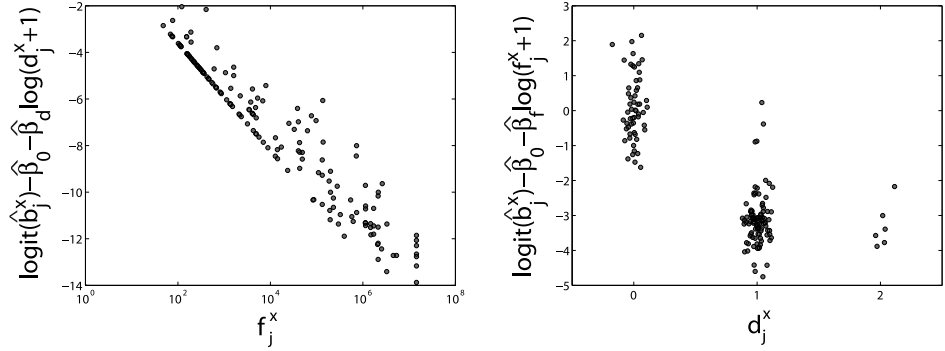}

\caption{Plots for all 52 root tweets of (left) $\operatorname
{logit}(\widehat{b}_j^x )-\widehat{\beta}_0-\widehat{\beta}_d\log
(d_j^x +1)$
versus $f_j^x $ and (right) $\operatorname{logit}(\widehat{b}_j^x
)-\widehat{\beta}_0-\widehat{\beta}_f\log(f_j^x +1)$
versus $d_j^x $. The values of $d_j^x $ are slightly perturbed in order
to improve
visibility of the data.}
\label{figregression}
\end{figure}

The ML estimates of the regression coefficients are $\widehat{\beta}
_0 = 1.99$, $\widehat{\beta}_f = -0.79$ and $\widehat{\beta}_d =
-4.31$ and
the $p$-values of the corresponding $t$-statistic are all significantly
less than
0.001,
indicating
a high significance for each coefficient. In Figure~\ref{figregression} we plot
$\operatorname{logit}(\widehat{b}_j^x )-\widehat{\beta}_0-\widehat
{\beta}_d\log(d_j^x +1)$ versus $f_j^x $ and
$\operatorname{logit}(\widehat{b}_j^x )-\widehat{\beta}_0-\widehat
{\beta}_f\log(f_j^x +1)$ versus $d_j^x $ in order
to show the isolated effect of each covariate.


The value for $\widehat{\beta}_f$ is negative, which is expected
given the way $\widehat{b}_j^x $ is
defined, but the value is greater than $-$1.
This result says that after controlling for
$d_j^x $, the average value of $M_j^x $ scales as $b_j^x f_j^x \sim
(f_j^x )^c$
for some $0<c<1$. Therefore,\vspace*{2pt} the number of retweets should grow
with the number of followers a user has, but at a decreasing rate.
The value for $\widehat{\beta}_d$ is also negative,
indicating that after controlling for $f_j^x $, a retweet is less likely
the farther we get from the root user.
Both of these findings are
in accordance with previous research on retweet graph
structure [\citet{reftwittermoon,refgoelRetweetDepth}] and
provide face validity to our results.

\section{Retweet model}\label{secmodel}
Our data analysis in Section~\ref{secdataOverview} provides us
with insights on the important properties of
the dynamics of retweeting and the structure of retweet graphs.
Based on these insights, we propose a Bayesian model for the evolution
of the
retweet graph of a root tweet.

\subsection{Generative model for retweet graph evolution}
Our generative model for the evolution of a retweet graph
can be described as follows.
We start with a single user $v^x_0$ who posts the root tweet $x$.
This user has a reaction time $S^x_0=0$ and $M^x_0$ children
who will eventually retweet $x$. Each child $v_j^x $ of $M^x_0$
generates a random reaction time
$S_j^x $ and an independent random number of children $M_j^x $. This
process repeats
recursively with every child generating a reaction time
and an independent random number of its own children.

%
\begin{figure}

\includegraphics{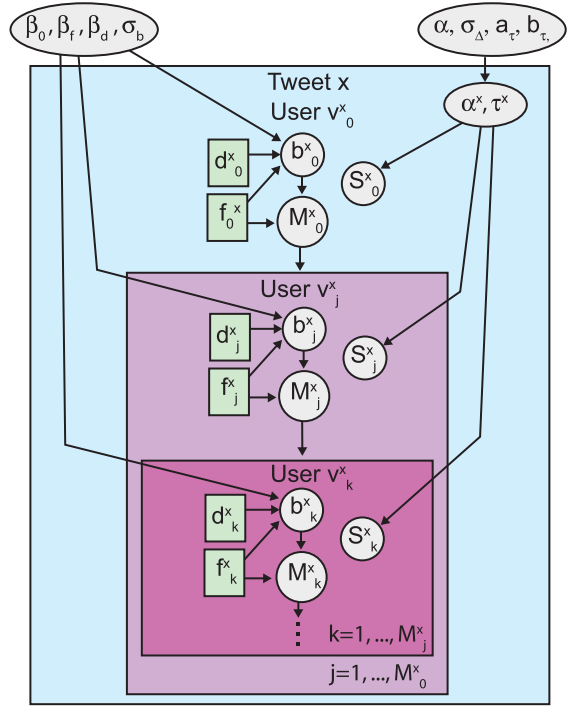}

\caption{Graphical model of the Bayesian log-normal-binomial
model for the evolution of retweet graphs. The plates denote
replication over tweets $x$
and users $v_j^x $. Nested plates denote retweets occurring
at larger depths from the root user. The process terminates when
all children which are leaves in the retweet
graph have $M_j^x =0$. Hyperpriors are omitted
for simplicity.}
\label{figmodel}
\end{figure}

The process terminates when all children which are leaves in the retweet
graph have $M_j^x =0$. As we show in our model specification of
Section~\ref{secmodelstructure}, the distribution of $M_j^x $
depends on the depth of the node and in
Section~\ref{secresults} we show that we typically learn that
$M_j^x $ is likely to be smaller for higher depth nodes.
The\vspace*{1.5pt} graphical
model of this generative model is shown in Figure~\ref{figmodel}.
In what follows, we specify the components of our
generative process by defining the conditional distributions of $S_j^x
$ and $M_j^x $.

\subsection{Log-normal model for reaction times}
From our exploratory analysis, we saw that a log-normal distribution
provided a reasonable fit for the reaction times. There was some variation
in the ML estimates of the log-normal parameters, $\alpha^x$ and $\tau
^x$, across tweets. Therefore,
we choose the following model for the reaction times.
For each root tweet $x$
we model $\log(S_j^x )$ as normal with a tweet specific mean
$\alpha^x$ and standard deviation $\tau^x$. We place
a normal prior on $\alpha^x$ and an inverse-gamma prior
on $(\tau^x)^2$, in accordance with standard hieararchical
Bayesian models [cf. \citet{refGelmanBook}]. In particular,
\begin{eqnarray}
\log\bigl(S_j^x \bigr)|\alpha^x,
\tau^x,M^x &\sim&\mathcal N\bigl(\alpha^x,
\bigl(\tau ^x\bigr)^2\bigr),\qquad j=1,\ldots,M^x,\label{eqlognormal}
\\
\alpha^x|\alpha,\sigma_\Delta& \sim&\mathcal N\bigl(\alpha,
\sigma _\Delta^2\bigr),
\\
\bigl(\tau^x\bigr)^2 &\sim& \operatorname{IG}(a_\tau,b_\tau).
\end{eqnarray}
To complete our hierarchical Bayesian specification and ameliorate issues
with hyperparameter sensitivity, we use the following hyperpriors:
\begin{eqnarray}
\alpha&\sim&\mathcal N\bigl(\mu_\alpha,\sigma_\alpha^2
\bigr),
\\
\sigma_\Delta^2 &\sim& \operatorname{IG}(a_\Delta,b_\Delta),
\\
\log(a_\tau)&\sim&\mathcal N\bigl(\mu_a,
\sigma_a^2\bigr),
\\
b_\tau&\sim&\operatorname{Gamma}(k_b,\theta_b),
\end{eqnarray}
and note that exact hyperparameter values, selected to be
uninformative, are provided
in Appendix \ref{appgibbs}.
The graphical model for the reaction time component
of the model is shown in
Figure~\ref{figmodel} (see\vspace*{2pt} node $S_j^x $ and all associated
connections)
and demonstrates the cross-tweet shrinkage that is
allowed by our model.


\subsection{Binomial model for retweet graph structure}\label{secmodelstructure}
As in our exploratory analysis, we assume independence of retweets
between the pool of potential retweeters, specifically assuming that
each follower of user $v_j^x $ retweets with probability $b_j^x $.
We saw initial evidence that the retweet probabilities $b_j^x $ showed
dependence
on the number of followers and depth of the user, $f_j^x $ and $d_j^x
$. Using\vspace*{2pt} this insight, we propose the following model
for the retweet graph structure:
\begin{eqnarray}
M_j^x |f_j^x,b_j^x &\sim&\operatorname{Bi} \bigl(f_j^x,b_j^x \bigr), \label{eqbinom}
\\
\operatorname{logit}\bigl(b_j^x \bigr)|
\mu_j^x,\sigma_b &\sim&\mathcal N \bigl(
\mu_j^x,\sigma_b^2 \bigr),
\end{eqnarray}
where we define
\begin{eqnarray}
\mu_j^x&=&\beta_0 + \beta_f\log
\bigl(f_j^x +1\bigr)+\beta_d\log
\bigl(d_j^x +1\bigr).
\end{eqnarray}
This model allows for the possibility of the
number of followers, $f_j^x $, and
the depth of the retweet from the root, $d_j^x $,
to influence the number of eventual retweeters.
The influence of the covariates, as determined by
$\beta_f$ and $\beta_d$, is shared across root tweets $x$.
As with the reaction time model, we put hyperpriors on these global
model parameters:
\begin{eqnarray}
\beta_0 &\sim&\mathcal N\bigl(\mu_{\beta_0},
\sigma_{\beta_0}^2\bigr),
\\
\beta_f &\sim&\mathcal N\bigl(\mu_{\beta_f},
\sigma_{\beta_f}^2\bigr),
\\
\beta_d &\sim&\mathcal N\bigl(\mu_{\beta_d},
\sigma_{\beta_d}^2\bigr),
\\
\sigma_b^2 &\sim& \operatorname{IG}(a_{\sigma_b},b_{\sigma_b}),
\end{eqnarray}
where we specify the specific (uninformative) hyperparameter values
in Appendix~\ref{appgibbs}. The combined model for reaction times
and the graph structure is shown in Figure~\ref{figmodel}.

\subsection{Likelihood function}
We now derive the likelihood function
for our retweet model. We partition our data set
into two types of tweets, \emph{training} tweets and \emph{prediction} tweets.
The training tweets are fully observed retweet graphs.
That is, we observe all reaction times ($S_j^x $) along with the final
degree ($M_j^x $) of each vertex in the retweet graph.
For the \emph{prediction} tweets, we observe the
retweet graph up to a time $t^x$ and
therefore only observe a fraction of the reaction times
and the current degree of each vertex which we denote by
$m_j^x (t^x)$.
We do not observe the $M_j^x $'s in a prediction\vspace*{2pt}
tweet\footnote{Except in the degenerate case where $m_j^x =f_j^x $, in which
case $M_j^x =m_j^x $.} and, therefore, we treat these as missing data.

First, we derive the likelihood of the
observations for a training tweet.
We define
the number of observed retweets for a training tweet $x$ as $m^x$.
The observed data for a training tweet are
$\mathbf S^x=\bigcup_{j=1}^{m^x}S_j^x $ and $\mathbf M^x=\bigcup_{j=0}^{m^x}M_j^x $.
Recall that in our model $\log(S_j^x )\sim\mathcal N(\alpha^x,(\tau
^x)^2)$ for $j=1,\ldots,m^x$.
Therefore,\vspace*{2pt}
if we define $\mathbf b^x=\bigcup_{j=0}^{m^x}b_j^x $,
the likelihood of the observations is given by
\begin{eqnarray}\label{eqtrain}
&& \mathbf P\bigl(\mathbf S^x,\mathbf M^x|
\alpha^x,\tau^x,\mathbf b^x,m^x
\bigr)\nonumber
\\
&&\qquad = P\bigl(M^x_0|b^x_0,F^x_0
\bigr)
\\
&&\quad\qquad{}\times \prod_{j=1}^{m^x}\frac{1}{\sqrt{2\pi}\tau^x} \exp
\biggl(-\frac{(\log(S_j^x )-\alpha^x)^2}{2(\tau^x)^2} \biggr)P\bigl(M_j^x
|b_j^x,f_j^x
\bigr),\nonumber
\end{eqnarray}
where $P(M_j^x |b_j^x,f_j^x )$ is given by the binomial of
equation~(\ref{eqbinom}). We note
that $S_j^x $ is not conditionally independent of $M_j^x $ because the total
number of $S_j^x $ that exist depend upon $M_j^x $
(which is an element in defining the observed $m^x$).


For\vspace*{1pt} the prediction tweets, we do not observe the $M_j^x $'s and so will need
to marginalize over them. Also, we observe only a subset of the
reaction times which comes from retweets that occur before time $t^x$.
Using the previous definitions of $\pi$ and $m^{x}(t^{x})$, the
observed data
for a prediction tweet are
$\mathbf S^x_{t^x}=\bigcup_{j=1}^{m^x(t^x)}S^x_{\pi(j)}$ and
$\mathbf m^x_{t^x}=\bigcup_{j=0}^{m^x(t^x)} m_{\pi(j)}^x(t^x)$.
First, we derive the conditional distribution
of the observations $\mathbf S^x_{t^x}$ and $\mathbf m^x_{t^x}$
conditional on
$\mathbf M^x_{t^x}=\bigcup_{j=0}^{m^x(t^x)} M^x_{\pi(j)}$, $\alpha
^x$ and $\tau^x$.
With\vspace*{1pt} this conditioning, the contribution to the probability from
each vertex $v^x_{\pi(j)}$ observed by time $t^x$ has three components:
\begin{longlist}[(3)]
\item[(1)] The log-normal likelihood of its observed reaction time
[equation~(\ref{eqlognormal})].
\item[(2)] The unobserved retweets of its children in the retweet graph.
That is, for each vertex $v^x_{\pi(j)}$ that retweets at time
$T^x_{\pi(j)}\leq t^x$,
we have $m^x_{\pi(j)}(t^x)$\vspace*{2pt} observed retweets by time $t$ and
$M^x_{\pi(j)}-m^x_{\pi(j)}(t^x)$ unobserved
retweets. Because
we are making the observations at time $t^x$, these $M^x_{\pi
(j)}-m^x_{\pi(j)}(t^x)$ reaction
times must be greater than $t^x-T^x_{\pi(j)}$. Therefore,
if we define the cumulative distribution function of $\mathcal N(\alpha
^x,(\tau^x)^2)$
as $F(\cdot|\alpha^x,\tau^x)$, the\break contribution to the conditional
distribution
is $(1-F(\log(t^x-T^x_{\pi(j)})|\break \alpha^x,\tau^x))^{M^x_{\pi
(j)}-m^x_{\pi(j)}(t^x)}$.
That is, $M^x_{\pi(j)}-m^x_{\pi(j)}(t^x)$ potential retweeters of
$v^x_{\pi(j)}$ have not done so yet (or we would have observed them by
time~$t^x$).
\item[(3)] A combinatorial term $M^x_{\pi(j)}\choose m^x_{\pi(j)}(t^x)$
which must be included because the unobserved retweets from the
children of
$v^x_{\pi(j)}$ could be any $M^x_{\pi(j)}-m^x_{\pi(j)}(t^x)$ of its
$M^x_{\pi(j)}$ children.
\end{longlist}
Putting these components together,
the likelihood of the prediction tweet
observations, conditional on the missing $M^x_{\pi(j)}$, is given by
%
\begin{eqnarray}\label{eqlikelihood}
&& \mathbf P \bigl(\mathbf S^x_{t^x},\mathbf
m^x_{t^x}|\alpha^x, \tau ^x, \mathbf
M^x_{t^x},m^x\bigl(t^x\bigr)
\bigr)\nonumber
\\[-1pt]
&&\qquad = \pmatrix{M^x_0 \cr m^x_0
\bigl(t^x\bigr)}
\bigl(1-F\bigl(\log\bigl(t^x-T^x_0\bigr)|
\alpha^x,\tau^x\bigr) \bigr)^{M^x_0-m^x_0(t^x)}
\nonumber\\[-8pt]\\[-8pt]
&&\quad\qquad{}\times \prod_{j=1}^{m^x(t^x)}\frac{1}{\sqrt{2\pi}\tau^x} \exp
\biggl(-\frac{(\log(S^x_{\pi(j)})-\alpha^x)^2}{2(\tau
^x)^2} \biggr)\pmatrix{M^x_{\pi(j)}\cr m^x_{\pi(j)}
\bigl(t^x\bigr)}\nonumber
\\[-1pt]
&&\hspace*{70pt}{}\times
\bigl(1-F\bigl(\log\bigl(t^x-T^x_{\pi(j)}\bigr)|
\alpha^x,\tau^x\bigr) \bigr)^{M^x_{\pi(j)}-m^x_{\pi(j)}(t^x)}.\nonumber
\end{eqnarray}
%
As can be seen from equaton (\ref{eqlikelihood}), for prediction
tweets $S_j^x $ and $M_j^x $ are not conditionally independent.
Because of this dependency we
can use temporal observations (retweet times) to
predict the final retweet graph structure (and
hence the final retweet count of the tweet).

To obtain the complete data likelihood, we simply multiply
equation~(\ref{eqlikelihood})
by $\mathbf P(M^x_{\pi(j)}|b^x_{\pi(j)},F^x_{\pi(j)})$
and sum\vspace*{2pt} over all possible values of $M^x_{\pi(j)}$. If we define
$\mathbf b^x_{t^x}=\bigcup_{j=0}^{m^x(t^x)}b^x_{\pi(j)}$,
then the marginal likelihood is
\begin{eqnarray}
&& \mathbf P \bigl(\mathbf S^x_{t^x},\mathbf
m^x_{t^x}|\alpha^x, \tau ^x,\mathbf
b^x_{t^x} \bigr)\nonumber
\\[-1pt]
&&\qquad = \sum_{M^x_0}\pmatrix{M^x_0
\cr m^x_0\bigl(t^x\bigr)} \bigl(1-F
\bigl(\log \bigl(t^x-T^x_0\bigr)|
\alpha^x,\tau^x\bigr) \bigr)^{M^x_0-m^x_0(t^x)}
\nonumber
\\[-1pt]
&&\quad\qquad{}\times \prod_{j=1}^{m^x(t^x)}\frac{1}{\sqrt{2\pi}\tau^x} \exp
\biggl(-\frac{(\log(S^x_{\pi(j)})-\alpha^x)^2}{2(\tau
^x)^2} \biggr)
\nonumber
\\[-1pt]
&&\quad\qquad{}\times \sum_{M^x_{\pi(j)}}P\bigl(M^x_{\pi(j)}|b^x_{\pi(j)},F^x_{\pi
(j)}\bigr)\pmatrix{M^x_{\pi(j)}\cr m^x_{\pi(j)}
\bigl(t^x\bigr)}\nonumber
\\[-1pt]
&&\hspace*{66pt}{}\times \bigl(1-F\bigl(\log\bigl(t^x-T^x_{\pi(j)}\bigr)|
\alpha^x,\tau^x\bigr) \bigr)^{M^x_{\pi(j)}-m^x_{\pi(j)}(t^x)}.\nonumber
\end{eqnarray}
Since this equation does not yield a closed form,
we rely on imputing the missing $M_j^x $ as described next in
Section~\ref{secposterior}.

\subsection{Posterior computations}\label{secposterior}
To summarize, our goal is to calculate a predictive
distribution for reaction times, and hence
the number of eventual retweets of a prediction tweet $x$, given
a set of observed (training) retweet paths and the partial history of
$x$ observed up to time $t^x$.
Recall that our model consists of three types of parameters.
First, there are the global parameters
$\Phi= \{\alpha,\sigma_\Delta,a_\tau,b_\tau,\beta_0,\beta
_f,\beta_d,\sigma_b \}$ which
are shared between tweets.
Second, there are tweet specific parameters
$\bolds\alpha=\bigcup_{x}\alpha^x$
and $\bolds\tau=\bigcup_{x}\tau^x$.
Third, there is a tweet and user specific parameter: the retweet probability
$b_j^x $. We define the set of all
retweet probabilities as $\mathbf b = \bigcup_{x,j}b_j^x $.

The final vertex degrees ($M_j^x $) are missing data for the
prediction tweets. We define $\mathcal P$ as the set of prediction
tweets and
$\mathcal T$ as the set of training tweets. We define the
set of unobserved $M_j^x $ for a tweet $x$ as
$\mathbf M^x = \bigcup_{j}M_j^x $. For the prediction tweets
we define $\mathbf M_{\mathcal P} =\bigcup_{x\in\mathcal P}\mathbf M^x$
and for the training tweets we define
$\mathbf M_{\mathcal T} =\bigcup_{x\in\mathcal T}\mathbf M^x$.
We define the set of observed reaction times for a tweet $x$ as
$\mathbf S^x = \bigcup_{j}S_j^x $ and the set of all reaction
times for both the training
and prediction tweets as
$\mathbf S = \bigcup_{x}\mathbf S^x$.
Using the conditional dependencies in our model as laid
out in Figure~\ref{figmodel}, the posterior distribution of
the model parameters and $\mathbf M_{\mathcal P}$
given $\mathbf S$ and $\mathbf M_{\mathcal T}$ can be written as
%
\begin{eqnarray}\label{eqposterior}
\mathbf P (\Phi,\bolds\alpha,\bolds\tau,\mathbf b,\mathbf
M_{\mathcal P}|\mathbf S, \mathbf M_{\mathcal T} )&\propto &  \mathbf P (\Phi
) \prod_{x}\mathbf P \bigl(\alpha ^x|
\alpha,\sigma_{\Delta} \bigr) \mathbf P \bigl(\tau^x|a_\tau,b_\tau
\bigr)\nonumber
\\
&&{}\times \prod_{x,j} \mathbf P \bigl(M_j^x
|b_j^x,f_j^x \bigr)\mathbf P
\bigl(b_j^x |\mu^x_j,
\sigma_b \bigr)
\nonumber\\[-8pt]\\[-8pt]
&&{}\times  \prod_{x\in\mathcal T} \mathbf P \bigl(\mathbf
S^x|\alpha^x,\tau ^x,\mathbf M^x
\bigr)
\nonumber
\\
&&{}\times  \prod_{x\in\mathcal P} \mathbf P \bigl(\mathbf
S^x,\mathbf m^x_{t^x}|\alpha^x,
\tau^x,\mathbf M^x \bigr).\nonumber
\end{eqnarray}
To examine our desired predictive distribution of $\mathbf M_{\mathcal
P}$, we sample
from equation~(\ref{eqposterior}) using an MCMC sampler which
involves sampling the model parameters in addition to $\mathbf
M_{\mathcal P}$.
The predictive distribution
is approximated by utilizing
samples of $\mathbf M_\mathcal P$. Also, despite
being potentially very high dimensional, the structure of the posterior
distribution lends itself to an efficient
parallelized implementation which can result in
significant speedup. The details of the stages of our
sampler along with the parallelized implementation are provided in the \hyperref[app]{Appendix}.



\section{Results}\label{secresults}
We partition
our data set into a set of $26$ training tweets $\mathcal T$
and a set of 26 prediction tweets $\mathcal P$.
We randomly divide the tweets such that the training and prediction
sets have similar
retweet count distributions. The specific
partition used can be found in the supplemental materials [\citet
{refsuppA}].
We aim to calculate the predictive distribution for $\mathbf
M_{\mathcal P}$
using a fixed observation fraction
of retweets for each prediction. For instance, for an observation
fraction of 10\%, we used as observations all data from the 26 training
tweets and the first 10\% of the total number of reaction times for
each of the 26
prediction tweets. Note that by fixing the observation \emph{fraction},
we are observing each prediction tweet up to a different time.
We use observation fractions ranging from 10\% to 100\%.
100~represents a fully in-sample analysis, and lower fractions are used
to understand how early on in a tweet's life predictions can be made.

\begin{figure}[b]

\includegraphics{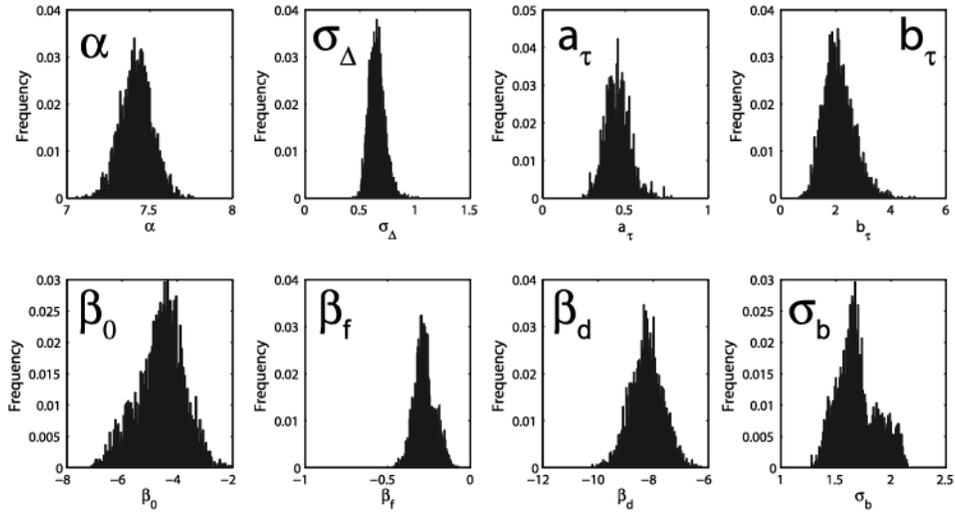}

\caption{Histograms of posterior samples of global
parameters with an observation fraction of~100\%.}\label{fighistPosterior}
\end{figure}

For each observation fraction, we generated posterior samples
using three independent MCMC chains with dispersed starting points run
for 3000 iterations and discarding a burn-in period of 1000 iterations.
Convergence of the MCMC sampler was assessed using the
Gelman--Rubin statistic [\citet{refgelman-rubin}]. A histogram of
the posterior samples
of the global parameters for an observation fraction of 100\% is shown
in Figure~\ref{fighistPosterior} and the corresponding posterior
means are shown in Table~\ref{tableparameters}.

%
%

%
\begin{table}
\tabcolsep=0pt
\tablewidth=180pt
\caption{Posterior means and standard deviations (s.d.)
for the global model parameters with
an observation fraction of 100\% (a fully in-sample analysis)}\label{tableparameters}
\begin{tabular*}{\tablewidth}{@{\extracolsep{\fill}}ld{2.8} @{}}
\hline
\textbf{Parameter} & \multicolumn{1}{@{}c@{}}{\textbf{Posterior mean (s.d.)}}\\
\hline
$\alpha$ & 7.42~(0.10) \\
$\sigma_\Delta$ & 0.65~(0.07) \\
$a_\tau$ & 0.45~(0.07) \\
$b_\tau$ & 2.11~(0.55) \\
$\sigma_b$ & 1.69~(0.18) \\
$\beta_0$ & -4.61~(0.85) \\
$\beta_f$ & -0.28~(0.06) \\
$\beta_d$ & -8.22~(0.59) \\
\hline
\end{tabular*}
\end{table}

We find that
the posterior mean of $\alpha$ is 7.42, which is comparable
to the mean of the ML estimates of $\alpha^x$ from
Section~\ref{secreaction} (7.31). Also, the 90\% posterior
credible interval
of the $\beta$ parameters do not contain 0, indicating
that these parameters are important to the predictive power
of our model and agree with our earlier analyses from Section~\ref{secgraph}.

In Section~\ref{secpredictions} we describe our prediction results
for the number
of eventual retweets, followed
by an analysis in Section~\ref{secstrawman} that looks at the impact
of the number of followers
($f_j^x $) and the depth of the retweeters ($d_j^x $) on our
predictions.

\subsection{Retweet prediction results}\label{secpredictions}
The predictions of our model for the total
number of retweets come from $M_j^x $,
the eventual number of retweets from retweeter $v_j^x $. For instance,\vspace*{2pt}
if at time $t^x$ we observe $m^x(t^x)$ retweets, our prediction
of the total number of retweets is given by
the predictive distribution of $\sum_{j=0}^{m^x(t^x)}M^x_{\pi(j)}$.
This serves as a step-ahead forecast of $M^x$. We discuss possibilities
to go beyond this step-ahead prediction in Section~\ref{secfutureRetweeters}.

%
\begin{figure}[t]

\includegraphics{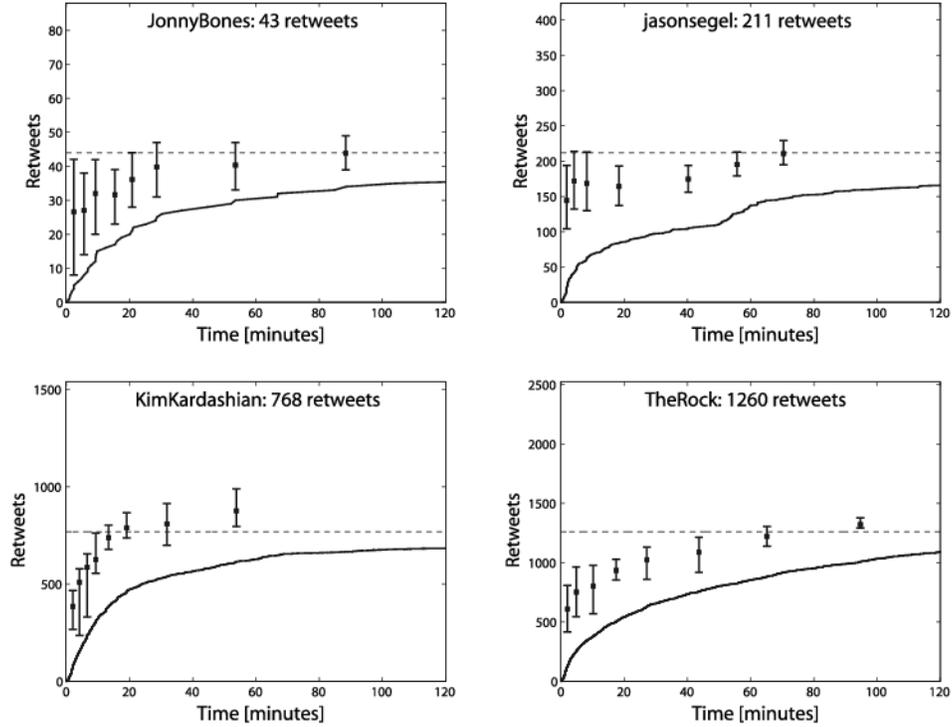}

\caption{Prediction of the total number of retweets for four
different root tweets. The solid line represents the number of observed
retweets versus time.
The solid square is the posterior median
of the predictive distribution for the total number of retweets based
on observations only up to that time point. The error bars correspond
to the 90\% credible intervals.
The horizontal dashed line is the final number of observed retweets $M^x$.
The root user and total number of retweets of each tweet are shown in
the plots.}\label{fighitzone}
\end{figure}

Our predictions are for observation fractions
ranging from 10\% to 100\%. The prediction results
for four different root tweets are shown
in Figure~\ref{fighitzone}. We plot the median
and $90\%$ posterior credible
intervals for the total number of retweets
for different observation
fractions. The predictions are plotted along
with the number of observed retweets versus
time. From these plots, it can be seen
qualitatively that the predictions made within a few
minutes for the eventual
number of retweets are relatively close to the
true value. We have found for all the prediction tweets
that the median time for the total number of retweets
to enter the $90\%$ posterior credible interval of the prediction
is 3 minutes.

%
\begin{figure}[t]

\includegraphics{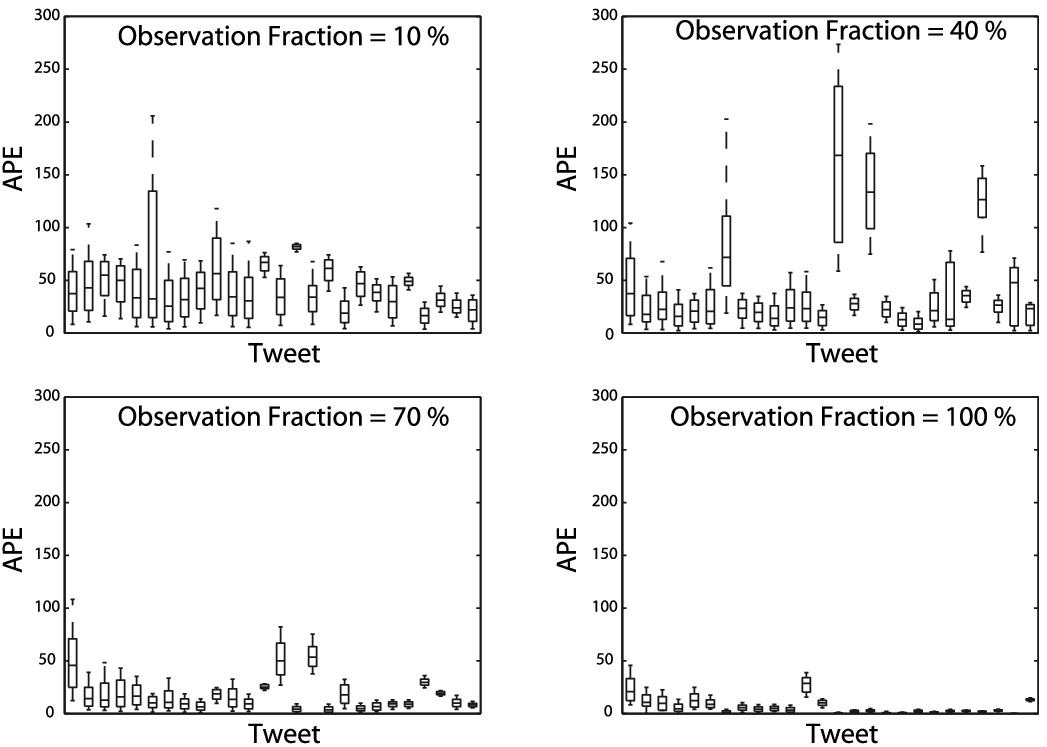}

\caption{Boxplots of prediction absolute percent error (APE)
for 26 prediction tweets. Each plot corresponds to a different
observation fraction of retweets.}\label{figboxplot}
\end{figure}

To better understand the model predictions at
the individual tweet level, we show boxplots
of the posterior distribution of the absolute
percent error (APE) for each prediction
tweet (using the posterior median as the prediction value)
for different observation fractions in Figure~\ref{figboxplot}.
The whiskers on the boxplots are the 90\%
posterior credible intervals. As can be seen, as
we increase the observation fraction,
the prediction error tends to decrease. There
are a few tweets which have exceptionally large
errors at a 40\% observation fraction. We discuss
these tweets in Section~\ref{secreaction2}.

\begin{figure}[b]

\includegraphics{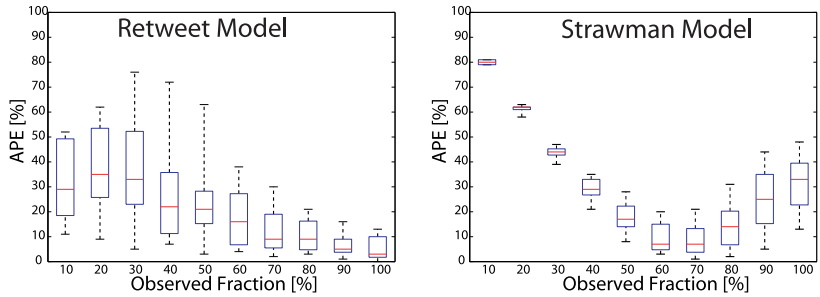}

\caption{Boxplots of the APE of the retweet model and
strawman model at different observation fractions.}
\label{figboxplotStrawman}
\end{figure}

We can aggregate these results across all prediction tweets
by looking at the APE of predictions
made using the posterior median as our prediction value.
We have found no significant relationship between the APE
of a prediction and the final number of retweets. For instance,
at 25\%, 50\% and 75\% observation
fractions the correlation between the APE and final number of retweets
is 0.14 ($p$-value 0.49), 0.14 ($p$-value 0.49) and 0.14 ($p$-value 0.49),
respectively.
In Figure~\ref{figboxplotStrawman} we show a boxplot of the APE
for all 26 prediction tweets versus observation fraction.

%
%

As can be seen, for our model the median APE (MAPE) is below 40\%
for observation fractions ranging from
10\% to 100\%. The average retweet time of the prediction
tweets at a 10\% observation fraction is 4.4 minutes. Therefore,
we see that using only a few minutes of observations, we can predict
with reasonable accuracy the total number of retweets given a small
fraction of observations. To check robustness, we have repeated
the predictions on 10 different random partitions of the tweets. We
have found for 10\% observation fraction the MAPE of each partition
was between 20\% and 36\%, with an average value of 28\%.

To get a sense
of how good the predictions are, consider the MAPE at 10\% and 100\%.
At 10\%, if one thought that there were no more retweets, the error
would be 90\%. Our model's median error is less than 40\%, which means
that the model predicts that the tweet will receive many more retweets.
At 90\%, if one thought the there were no more retweets, the error
would be 10\%. Our model's median error is less than 10\%, which means
that the model predicts that the tweet is almost done spreading.
Therefore, we see that
our model can predict if a tweet has a significant amount of (retweet)
life left
or if it is near its end.

\subsection{Comparison with benchmark models}\label{secbenchmark}
We next compare our model with three different benchmark models. First,
we consider a linear regression model that uses no temporal information
and only the follower count of the root user (source tweeter). Second,
we consider the regression
model of \citet{refszabo} which uses only the current retweet
count. Finally,
we consider a dynamic Poisson model with exponentially decaying rate
based on
the work of \citet{refagarwal}. We will see that our model outperforms
each of these approaches.

The linear regression model is as follows:
\begin{eqnarray}
\log\bigl(M^x\bigr) & =&\beta_0 + \beta_1
\log\bigl(f^x_0\bigr)+\varepsilon^x,
\end{eqnarray}
where $\varepsilon^x$ is a zero mean, normally distributed error term.
This model only uses the root users' follower count to predict the final
retweet count, but no information about the retweet times or followers
and depth
of retweeters.

The regression model of \citet{refszabo} for the final retweet count
is
\begin{eqnarray}
\log\bigl(M^x\bigr) & =&\beta(t) + \log\bigl(m^x(t)
\bigr)+\varepsilon^x,
\end{eqnarray}
where $\varepsilon^x$ is a zero mean, normally distributed error term.
Here the final retweet count is modeled as a log-linear function of the
current retweet
log count at time $t$, where the intercept $\beta(t)$ is time varying. Since
$m^x(t)$ approaches $M^x(t)$ as
$t$ goes to infinity, we also expect $\beta(t)$ to
approach zero in this model.

For the dynamic Poisson model with exponentially decaying rate, we bin
time into 5 minute intervals indexed by $k=0,1,2,\ldots.$ The number of
retweets in the $k$th bin is a
Poisson random variable with rate $\lambda\delta^k$. Here $\lambda$
is the initial
retweet rate, and $\delta$ describes the exponential decay of the rate.

%
\begin{figure}

\includegraphics{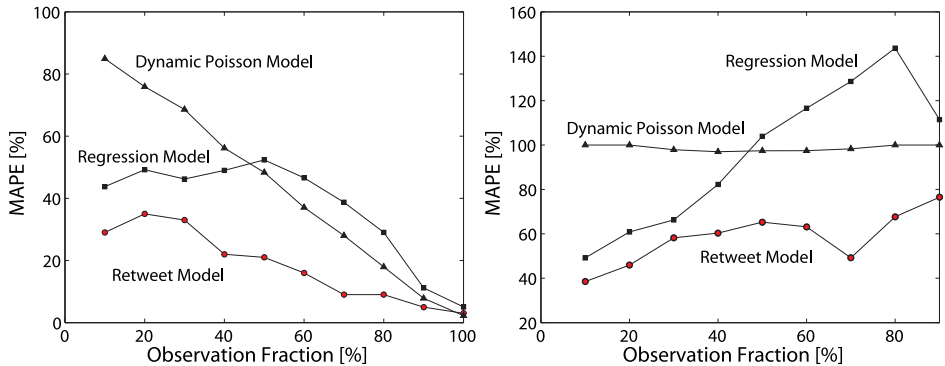}

\caption{Plots of the median absolute percentage error (MAPE) for the
total retweet count (left) and remaining retweet count (right)
versus observation fraction of retweets for 26 root tweets. The
three curves are the MAPE for the retweet model, the linear regression
model of \citet{refszabo} and the dynamic Poisson model with
exponentially
decaying rate.}\label{figbenchmark}
\end{figure}

We perform ML estimation of these models on the training tweets, and
then predict on the
prediction tweets. For the linear regression model which only uses the
follower count,
the MAPE is 65\%. This is much higher than our model that is able to use
observations of retweet times. For the other two models which utilize
retweet times,
we plot their MAPE
in Figure~\ref{figbenchmark}. We plot the MAPE of
both the final retweet count and also the remaining retweet count
(so that the maximum possible MAPE${} ={}$100\%).
For each type of MAPE, we can see that our retweet model outperforms
the other models.

\subsection{Comparison with nested models: Impact of $f_j^x$ and $d_j^x$}\label{secstrawman}
\begin{figure}

\includegraphics{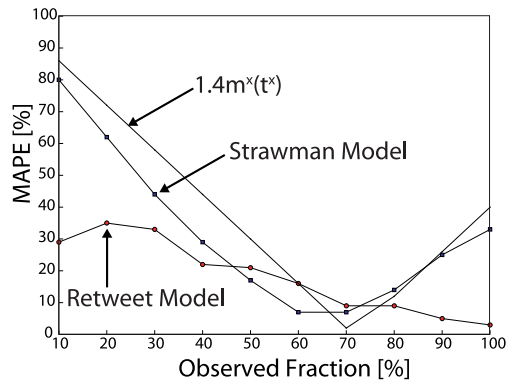}

\caption{Plot of the median absolute percentage error (MAPE)
versus observation fraction of retweets for 26 root tweets. The
three curves are the MAPE for the retweet model, a strawman
model which ignores $f_j^x $ and $d_j^x $, and a naive model
which always predicts $1.4m^x(t^x)$.}\label{figmape}
\end{figure}

To show the importance of $f_j^x $ and $d_j^x $ to our
retweet model, we compare to a strawman model which
ignores these covariates. The strawman model assumes
that $M_j^x $ comes from a Poisson distribution (not binomial as before
since $f_j^x $ is unknown)
with global
rate $\lambda$. We keep
the reaction time component of the retweet model the same.
We put an uninformative gamma prior on $\lambda$ with shape
and scale parameters 1 and~500, respectively.
We use the median
of the predictive distribution as a point estimate
of the number of retweets in comparing our model's performance
to that of the strawman.
In Figure~\ref{figboxplotStrawman} we show boxplots
for the absolute percent error (APE)
of the two models' predictions
for all of the prediction tweets versus
the observation fraction. For an observation
fraction of 10\% (where predictions
are most useful) the
error of the strawman model is very high (MAPE${}={}$80\%)
compared to our model (MAPE${}={}$29\%). Also, while
our model's error tends to decrease as more retweets are
observed, the strawman model's error decreases to a point and then
increases again. The strawman model's prediction for the total
number of retweets is essentially a constant multiplied by the
number of observed retweets. To make this more evident, in Figure~\ref{figmape} we plot
the MAPE versus observation fraction for both models and a
naive model which predicts $1.4m^x(t^x)$
for the eventual number of retweets. The factor of $1.4$ was
chosen to make the minimum MAPE of the naive model occur
at the same observation fraction as the strawman model.
As can be seen, the error of the strawman is very similar to the naive
model.

%

To assess the overall fit of the two models, we compare
their average log-likelihood (LL) and deviance information
criterion (DIC) [\citet{refdic}] for an observation fraction of
100\% in Table~\ref{tabledic}.
Models which fit better have larger values for the LL and
smaller values for the DIC.
As can be seen from Table~\ref{tabledic}, our model
has a significantly better fit than the strawman model. This analysis
demonstrates that $f_j^x $ (user information)
and $d_j^x $ (retweet\vspace*{2pt} graph structure) are important elements
for predicting retweets accurately.

\begin{table}[b]
\tabcolsep=0pt
\tablewidth=200pt
\caption{Average log-likelihood (LL) and deviance
information criterion (DIC) for a 100\% observation
fraction for the full retweet model and a nested strawman model}\label{tabledic} 
\begin{tabular*}{\tablewidth}{@{\extracolsep{\fill}}@{}lcc@{}} 
\hline
& \textbf{Retweet model} &\textbf{Strawman model}\\
\hline
LL & $-$38,860 &$-$103,907\\
DIC & \phantom{$-$}83,848 & \phantom{$-$}208,026\\
\hline
\end{tabular*}
\end{table}
%

\section{Model extension opportunities}\label{secextensions}
We next discuss various extensions to our
retweet model. We first discuss improving
our predictions using future potential retweeters.
Then we discuss evidence in our data which suggests
possible extensions to our reaction
time model. Finally, we discuss
the incorporation of side information for the tweets.

\subsection{Distribution over future potential retweeters}\label{secfutureRetweeters}
Our current predictions are based on eventual retweets from
existing users in the observed retweet graphs and do
not take into account retweets of future retweeters who
have not yet been observed. We can think of this prediction
as a step-ahead forecast of the total eventual number
of retweeters. In practice, it quickly provides a good
estimate since most retweet graphs have low depth and retweets occur
quickly. However,
one could extend our prediction to account for the eventual retweets
from users who have not yet been observed, in particular, by
integrating over our uncertainty.
This type of prediction
would require greater knowledge of the structure of the underlying
follower graph. For instance, if a user has a follower
with a large number of followers, this user may receive
a large number of retweets due to a retweet from
this follower. Therefore, incorporation of unobserved
retweeters could potentially improve
our predictions, but would require obtaining more data
on the follower graph. Note, however, that under the
(experimentally validated) assumption that the probability
of retweeting decreases with depth, the sensitivity of our
predictions to inaccuracies of future retweeter information may be
minimal.

\subsection{Reaction time modeling}\label{secreaction2} As seen in Figure~\ref{figboxplot} (top right),
at an observation fraction of 40\%
there are four different
tweets with very large errors compared
to the other tweets. We looked at these
tweets more closely to try to understand
the source of this error. The number
of retweets for these tweets ranged
from 73 to~608. What these tweets had in
common was the fact that the number of retweets
increased very rapidly at first, and then slowed
down considerably. This behavior deviated from
the log-normal reaction time model.
%
%
\begin{figure}[b]

\includegraphics{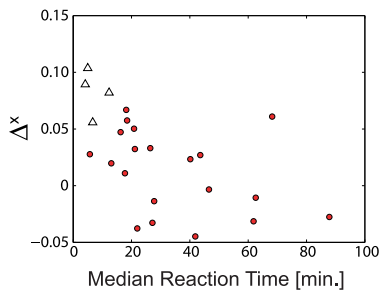}

\caption{Plot of median reaction time versus
$\Delta^x$ for the prediction tweets.
The triangle points are the tweets with large prediction errors
at 40\% observation from Figure \protect\ref{figboxplot}.}
\label{figerror}
\end{figure}
If the reaction
times were log-normal, then their logarithms
would be normally distributed and the difference between the
median and mean
of their logarithms would be zero. Any deviation
of this difference from zero can be viewed as a deviation
from log-normality. We define $\Delta^x$ as
this difference normalized by the
median of the logarithm of the reaction times:
\[
\Delta^x = \frac{\mathrm{mean}(\log(S_j^x ))-\mathrm{median}(\log
(S_j^x))}{\mathrm{median}(\log(S_j^x ))}.
\]

To show the similarities of the four high error tweets,
in Figure~\ref{figerror} we plot $\Delta^x$ versus the
median reaction time for each
prediction tweet. The four triangles in the plot are the tweets
with the large errors. As can be seen, these tweets have a short
median reaction time along with a large value for $\Delta^x$. Therefore,
it seems that these tweets have reaction times that are not well
modeled by the log-normal distribution, which leads to the
larger prediction errors. It is an interesting area of future
research to try and understand what properties of these tweets
and the users who posted them cause this type of retweeting behavior
and why the reaction times are not well modeled by the log-normal
distribution.

\subsection{Incorporation of side information}
Our model relied primarily on the timing information
of retweets, depth in the retweet graph and number
of followers for predictions. However, there are other
types of side information that we could incorporate
which may potentially improve the accuracy of the
predictions. One type of side information is the time of day. It may
be that the retweet behavior of a tweet depends
upon the time it was posted. Another type of side information
is the content of
the tweet. For instance, retweet behavior may depend
upon the topic of the tweet, and whether or not that
topic is a currently trending topic in Twitter.
These types of side information can be readily
incorporated into our modeling framework as covariates for the
parameters such as $\alpha^x$ and $b_j^x $.

\section{Conclusion}
We have presented a model for
retweet dynamics in Twitter. Our
Bayesian approach allowed us to provide predictions
for the total number of retweets, along with posterior credible
intervals for the predictions. The predictions had a MAPE of
less than 40\% when at least 10\% of the total number of
retweets were observed. For most tweets, this translated
to an average error less than 40\% within 5 minutes of the
tweet being posted.

We have shown that given
the size of the retweeter network and depth from the source tweet, we
are able to predict the number
of potential viewers of a tweet.
The level of accuracy in our predictions allows us
to consider using this model for different applications.
For example, it can be used to turn tweets into a potential source
of impressions for display ads. Because tweets are typically only
actively retweeted for a few hours, the early
predictions our model provides are key to detecting
a popular tweet before it receives a large amount of
retweets. Also, the similarity of
the manner by which people spread
content in social networks suggest that this model can be used for
other social networks such as Facebook. Therefore,
our model's early predictions could create a whole
new source of impressions for online advertising on dynamic
social network content with a finite ``lifetime.''

Finally, because this model is
for a single tweet, it can be used as the foundation
for a more general model for the spread of
broader ideas which involve multiple tweets
from multiple users. Our model can easily
be parallelized to analyze very large collections
of tweets. With a model for the spread of ideas, we could
develop a better understanding of how memes and trends
spread and potentially predict the speed
and magnitude of their popularity.

\begin{appendix}\label{app}
\section{Details of MCMC sampler}\label{appgibbs}

We use a Metropolis-within-Gibbs scheme to sample
from the posterior distribution of the model
parameters. We define the set of model parameters as
$\Theta= \{\Phi,\mathbf b, \bolds\alpha^x,\bolds
\tau^x, \mathbf M_{\mathcal P} \}$ and for
any parameter $\gamma\in\Theta$, we define the set
of parameters excluding $\gamma$ as
$\Theta_{-\gamma}$. We also define
the set of observed reaction times as $\mathbf S$.
For our MCMC sampler, we must
sample from the conditional distribution
$\mathbf P (\gamma|\mathbf S,\mathbf M_{\mathcal T},\Theta
_{-\gamma} )$
for each model parameter. We will now
derive these conditional distributions
and show how to sample from them.

\subsection{Retweet graph structure parameters}

\subsubsection*{Hyperparameters $\beta_0$, $\beta_F$, $\beta_d$, $\sigma_b^2$}
The prior distributions for $\beta_0$, $\beta_F$ and $\beta_d$ are
normal with mean 0 and standard deviation $\sigma_\beta=100$.
It can be shown that the joint conditional
distribution of $(\beta_0,\beta_F,\beta_d)$
is multivariate normal with mean $\bolds\mu$ and covariance
matrix $\mathbf C$. Because
of this, we can directly sample the $\beta$'s in
a Gibbs step. We simply need to determine $\bolds\mu$
and $\mathbf C$. To do this, first we let $N$ be the total
number of observed reaction times for all training
and prediction tweets.
To express the mean and covariance
of the conditional distribution, it is helpful to define the following
variables:
\begin{eqnarray*}
N_1 &=& N+\sigma^2_b\sigma_{\beta}^{-2},
\qquad E = \sum
_{x,j}\log \bigl(f_j^x
+1\bigr)\log\bigl(d_j^x +1\bigr),
\\
D &=& \sum
_{x,j}\log\bigl(d_j^x
+1\bigr),\qquad D_2 = \sum
_{x,j}\log
^2\bigl(d_j^x +1\bigr)+\sigma^2_b
\sigma_{\beta}^{-2},
\\
F &=& \sum
_{x,j}\log\bigl(f_j^x
+1\bigr),\qquad F_2 = \sum
_{x,j}\log
^2\bigl(f_j^x +1\bigr)+\sigma^2_b
\sigma_{\beta}^{-2},
\\
Y_0 &=& \sum
_{x,j}\log\bigl(b_j^x
+1\bigr),\qquad Y_F = \sum
_{x,j}\log
\bigl(b_j^x +1\bigr)\log\bigl(f_j^x
+1\bigr),
\\
Y_d &=& \sum
_{x,j}\log^2
\bigl(b_j^x +1\bigr)\log\bigl(d_j^x
+1\bigr)+\sigma ^2_b\sigma_{\beta}^{-2}.
\end{eqnarray*}
Then the covariance matrix of the conditional distribution
is given by
\begin{eqnarray*}
\mathbf C &=& \sigma_b^2\lleft[ %
\matrix{
N_1 & F & D
\cr
F & F_2 & E
\cr
D & E &D_2}
 \rright]^{-1}
\end{eqnarray*}
and its mean is given by
\begin{eqnarray*}
\bolds\mu&=& \lleft[ %
\matrix{ N_1 & F & D
\cr
F &
F_2 & E
\cr
D & E &D_2} %
 \rright]^{-1}
\lleft[ %
\matrix{ Y_0
\cr
Y_F
\cr
Y_d} %
 \rright].
\end{eqnarray*}

The prior distribution of $\sigma_b^2$ is inverse-gamma
with shape and scale parameters $a_{\sigma_b}=0.5$ and $b_{\sigma
_b}=0.5$, respectively.
We can directly sample from the conditional distribution
for $\sigma_b^2$ because it is inverse-gamma with shape parameter
$a'_{\sigma_b}$ and scale parameter $b'_{\sigma_b}$ given by
\begin{eqnarray*}
a'_{\sigma_b}&=&a_{\sigma_b}+\frac{N}{2},
\\
b'_{\sigma_b}&=&b_{\sigma_b}+\frac{1}{2}\sum
_{x,j} \bigl(\operatorname{logit}\bigl(b_j^x
\bigr)-\mu^x_j \bigr)^2,
\end{eqnarray*}
where $\mu_j^x = \beta_0+\beta_F\log(f_j^x +1)+\beta_d\log(d_j^x +1)$.

\subsubsection*{Parameters $b_j^x$}
The conditional distribution of $b_j^x $ is given by
\begin{eqnarray}
\mathbf P \bigl(b_j^x |\mathbf S,\mathbf
M_{\mathcal T},\Theta _{-b_j^x } \bigr) &\propto&\mathbf P
\bigl(M_j^x |b_j^x \bigr)\mathbf
P \bigl(b_j^x |\beta_0,
\beta_F,\beta_d,\sigma_b \bigr)
\nonumber
\\
&\propto&\bigl(b_j^x\bigr)^{M_j^x }
\bigl(1-b_j^x \bigr)^{f_j^x -M_j^x } \exp \biggl(-
\frac{ (\operatorname{logit}(b_j^x )-\mu_j^x
)^2}{2\sigma_b^2} \biggr).
\nonumber
\end{eqnarray}
To sample from this conditional distribution, we use a Metropolis--Hastings
step with the proposal value for $\operatorname{logit} (b_j^x
 )$ drawn from
a normal
distribution with mean $\mu_j^x$ and standard deviation $\sigma_{b}$.

\subsubsection*{Missing $M_j^x$}
The conditional distribution for $M_j^x $ is
\begin{eqnarray*}
\mathbf P \bigl(M_j^x |\mathbf S,\mathbf
M_{\mathcal T},\Theta_{-M_j^x }\bigr)& \propto& \pmatrix{M_j^x
\cr m_j^x } \bigl(1-F\bigl(\log\bigl(t-S^x_j
\bigr)|\alpha^x,\tau\bigr) \bigr)^{M_j^x -m_j^x }
\\
&&{}\times \pmatrix{f_j^x \cr M_j^x }
\bigl(b_j^x \bigr)^{M_j^x } \bigl(1-b_j^x
\bigr)^{f_j^x -M_j^x }\mathbf1 \bigl\{M_j^x \geq
m_j^x \bigr\}.
\end{eqnarray*}
We generate samples from this conditional distribution using a
Metropolis--Hastings
step with the proposal for $M_j^x $ drawn from a binomial distribution
$\operatorname{Bi}(f_j^x,b_j^x )$.

\subsection{Retweet time parameters}

\subsubsection*{Hyperparameters $\alpha$, $\sigma_\Delta^2$, $a_{\tau}$, $b_{\tau}$}
We utilized an extremely diffuse prior distribution for $\alpha$ that
is normal
with mean $0$ and standard deviation $\sigma_\alpha=100$.
The conditional distribution of $\alpha$ is again normal with mean
$\mu_{\alpha}'$ and variance $\sigma'^{2}_\alpha$, so it can
be directly sampled. If we define the total number of root tweets
(training and prediction) as $N_t$, then the mean and variance are
\begin{eqnarray*}
\mu'_{\alpha} & =& \bigl(N_t+
\sigma_\Delta^2\sigma_\alpha ^{-2}
\bigr)^{-1}\sum
_{x}\alpha^x,
\\
\sigma_{\alpha}'^2& =& \bigl(N_t+
\sigma_\Delta^2\sigma_\alpha ^{-2}
\bigr)^{-1}\sigma_\Delta^2.
\end{eqnarray*}

The prior distribution of $\sigma_\Delta^2$ is inverse-gamma
with shape and scale parameters $a_{\sigma_\Delta}=0.5$ and
$b_{\sigma_\Delta}=0.5$, respectively.
We can directly sample from the conditional distribution
for $\sigma_\Delta^2$ because it is again inverse-gamma with shape parameter
$a'_{\sigma_\Delta}$ and scale parameter $b'_{\sigma_\Delta}$ given by
\begin{eqnarray*}
a'_{\sigma_\Delta}&=&a_{\sigma_\Delta}+\frac{N_t}{2},
\\
b'_{\sigma_\Delta}&=&b_{\sigma_\Delta}+\frac{1}{2}\sum
_{x} \bigl(\alpha^x-\alpha \bigr)^2.
\end{eqnarray*}

The prior distribution of $\log(a_\tau)$ is normal
with mean $\mu_a=0$ and standard deviation $\sigma_a=10$.
The conditional distribution of $a_\tau$ is given
by
\begin{eqnarray*}
\mathbf P (a_\tau|\mathbf S,\mathbf M_{\mathcal T},\Theta
_{-\alpha^x} )&\propto &\mathbf P (a_\tau )\prod
_{x=1}^{N_t}\mathbf P \bigl(\tau^x|a_\tau,b_\tau
\bigr)
\\
&=&\exp \biggl(-\frac{\log^2(a_\tau)}{2\sigma_a^2} \biggr) \prod_{x=1}^{N_t}
\frac{b_\tau^{a_\tau}}{\Gamma(a_\tau)} \bigl(\tau ^x\bigr)^{-a_\tau}.
\end{eqnarray*}

To sample from this conditional distribution, we use a random walk
Metropolis--Hastings step.
That is, if we define the $i$th sample of $a_\tau$ as
$a_{\tau,i}$, the proposal for the $(i+1)$ sample is drawn from a normal
distribution with mean $a_{\tau,i}$ and standard deviation 0.2, where
0.2 is chosen to balance the acceptance rate with step size.

The prior distribution of $b_\tau$ is gamma
with shape parameter $k_b=1$ and scale parameter $\theta_b=500$.
We can sample directly from the conditional distribution
of $b_\tau$ because it is gamma with shape parameter
$k'_b$ and scale parameter $\theta'_b$ given by
\begin{eqnarray*}
k'_{b}&=&k_b+N_ta_\tau,
\\
\theta'_{b}&=& \biggl(\theta_b^{-1}+
\sum
_{j} \bigl(\tau ^x \bigr)^{-1}
\biggr)^{-1}.
\end{eqnarray*}

%
\subsubsection*{Parameters $\alpha^x$, $\tau^x$}
The conditional
distribution of $\alpha^x$ depends upon
whether the root tweet is in the training or
prediction set. For training tweets, the conditional
distribution of $\alpha^x$ is normal with mean
$\mu_{\alpha^x}$ and variance $\sigma_{\alpha_x}^2$ with
\begin{eqnarray*}
\mu_{\alpha^x} & =& \bigl(M^x+\tau^2
\sigma_\Delta^{-2} \bigr)^{-1}\sum
 _{j=1}^{N_t}
\log\bigl(S_j^x \bigr),
\\
\sigma_{\alpha^x}^2& =& \bigl(M^x+\tau^2
\sigma_\Delta^{-2} \bigr)^{-1}\tau^2.
\end{eqnarray*}
For a prediction tweet with $n$ observed retweets, the conditional
distribution of $\alpha^x$ is given
by
\begin{eqnarray*}
&& \mathbf P \bigl(\alpha^x|\mathbf S,\mathbf M_{\mathcal T},\Theta
_{-\alpha^x} \bigr)
\\
&&\qquad \propto \exp \biggl(\frac{(\alpha^x-\alpha)^2}{2\sigma_\Delta^2} \biggr)
\\
&&\quad\qquad{}\times \prod
_{j=0}^{n-1} \exp \biggl(-\frac{(\log(T^x_{j+1})-\alpha^x)^2}{2\tau^2}
\biggr)
\bigl(1-F\bigl(\log\bigl(t-S^x_j
\bigr)|\alpha^x,\tau\bigr) \bigr)^{M_j^x -m_j^x }.
\end{eqnarray*}
To sample from this conditional distribution, we use a random walk
Metro\-polis--Hastings step. We define the $i$th sample of $\alpha^x$ as
$\alpha^x_i$, and the proposal for the $(i+1)$ sample is drawn from a normal
distribution with mean $\alpha^x_i$ and standard deviation~0.2, where
0.2 is chosen to balance the acceptance rate with step size.

The prior distribution of $(\tau^x)^2$ is inverse-gamma
with shape and scale parameters $a_\tau$ and $b_\tau$, respectively.
We denote the inverse-gamma density function by $\operatorname{IG}(\cdot|a_\tau,b_\tau)$.
The conditional
distribution of $(\tau^x)^2$ can be written as
\begin{eqnarray*}
&& \mathbf P \bigl(\bigl(\tau^x\bigr)^2|\mathbf S,\mathbf
M_{\mathcal T},\Theta_{-\tau} \bigr)
\\
&&\qquad \propto \operatorname{IG}\bigl(\bigl(
\tau^x\bigr)^2|a'_\tau,b'_\tau
\bigr)\prod
_{x\in\mathcal
P} \bigl(1-F\bigl(\log\bigl(t-S^x_j
\bigr)|\alpha^x,\tau\bigr) \bigr)^{M_j^x -m_j^x },
\end{eqnarray*}
where the parameters of the inverse-gamma density function above are
\begin{eqnarray*}
a'_\tau&=& a_\tau+\frac{m^x(t)}{2},
\\
b'_\tau&=& b_\tau+ \frac{1}{2}\sum
_{j=1}^{m^x(t)} \bigl(\log \bigl(S_j^x
\bigr)-\alpha^x \bigr)^2.
\end{eqnarray*}
For training tweets, $M_j^x =m_j^x $, so the conditional distribution
is inverse-gamma
and we can sample $\tau^x$ directly. For prediction tweets, we must
use a Metropolis--Hastings step with the proposal value for $(\tau^x)^2$
drawn from an inverse-gamma distribution
with shape and scale parameters $a'_\tau$ and $b'_\tau$, respectively.

\section{Distributed implementation of MCMC~sampler}\label{appalgo}
The MCMC sampler lends itself naturally to distributed computation.
The variables to be sampled are global (shared) and local (tweet/user
specific). The main computational
burden comes from the local random\vadjust{\goodbreak} variables, of which there can be
thousands or millions, depending on the size of the observations. However,
the steps for sampling many of these local variables can be done
simultaneously, which
can result in a considerable speedup.

There are two random variables associated with
each tweet/user pair $(x,j)\dvtx b_j^x $ and $M_j^x $. The only local
variable the sampling
step of $b_j^x $ depends on is $M_j^x $. For sampling $M_j^x $, the only
local variables needed are $b_j^x $, $\alpha^x$ and $\tau^x$.
Therefore, the sampling steps of $b_j^x $ and $M_j^x $ must be\vspace*{1pt}
done sequentially. However,\vspace*{1pt} this sequence of steps can
be done in parallel across all tweet/user pairs $(x,j)$.

There are two random variables associated solely with
each tweet $x$: $\alpha^x$ and $\tau^x$. The sampling
of $\alpha^x$ needs the values of $\tau^x$ and all $M_j^x $
associated\vspace*{1pt} with tweet $x$. Similarly, the sampling
of $\tau^x$ depends on the values of $\alpha^x$ and\vspace*{1pt} all $M_j^x $
associated with tweet $x$. Therefore,
the sampling steps of $\alpha^x$ and $\tau^x$ must be done sequentially,
but this can be done in parallel
across all tweets~$x$.

Putting all this together, we obtain the following distributed implementation
of the MCMC sampler to generate a sample from the full posterior distribution.
First, sequentially sample the global parameters $\Phi$.
Second, sequentially sample
the parameters $\alpha^x$ and $\tau^x$ for a tweet $x$, but simultaneously
for all tweets. Third, sequentially sample the parameters $b_j^x $ and
$M_j^x $
for all tweet/user pairs $(x,j)$, but simultaneously
for all tweet/user pairs. This results in a classic
data parallel setup that can be efficiently implemented using
frameworks such as MapReduce.
\end{appendix}

\begin{supplement}[id=suppA]
\sname{Supplement}
\stitle{Retweet time series data}
\slink[doi]{10.1214/14-AOAS741SUPP} 
\sdatatype{.zip}
\sfilename{AOAS741\_supp.zip}
\sdescription{These files contain the data of the retweet time series
for the root tweets studied in this paper.
They also include the files which contain the different partitions
of the tweets into training and prediction
sets used for the analysis in this paper.}
\end{supplement}


%

\printaddresses
\end{document}